\documentclass[useAMS,usenatbib]{mn2e}

\usepackage[latin1]{inputenc}  
\usepackage{graphicx}
\usepackage{txfonts}

\title[Photodestruction of the benzene molecule]{Dissociation of the benzene molecule by UV and soft X-rays in circumstellar environment}
\author[H. M. Boechat-Roberty]{H. M. Boechat-Roberty$^{1}$\thanks{E-mail:
heloisa@ov.ufrj.br}, R. Neves$^{1}$, S. Pilling$^{2}$, A. F. Lago$^{3}$ \and G.G. B. de Souza$^4$ \\
\\
$^{1}$Observatório do Valongo, Universidade Federal do Rio de
Janeiro - UFRJ, Ladeira Pedro Antônio 43, CEP 20080-090, Rio de
Janeiro,RJ, Brazil.\\
$^{2}$Pontifícia Universidade Católica do Rio de Janeiro, Marquês
de S. Vicente 255 CEP22543-970, Rio de Jeneiro, Brazil. \\
$^{3}$Universidade Federal do ABC, Rua Santa Adélia, 166, Santo André, São Paulo, Brazil. \\
 $^{4}$Instituto de Química, Universidade Federal do Rio de Janeiro - UFRJ, Ilha do
Fundão, CEP 21949-900, Rio de Janeiro, RJ, Brazil.}
\begin{document}

\date{Received / Accepted}

\pagerange{\pageref{firstpage}--\pageref{lastpage}} \pubyear{2005}

\maketitle

\label{firstpage}


\begin{abstract} 
Benzene molecules, present in the proto-planetary nebula CRL 618,
are ionized and dissociated by UV and X-ray photons originated from
the hot central star and by its fast wind. Ionic species and free
radicals produced by these processes can lead to the formation of
new organic molecules. The aim of this work is to study the
photoionization and photodissociation processes of the benzene
molecule, using synchrotron radiation and time of flight mass
spectrometry.  Mass spectra were recorded at different energies
corresponding to the vacuum ultraviolet (21.21 eV) and soft X-ray
(282 - 310 eV) spectral regions. The production of ions from the
benzene dissociative photoionization is here quantified, indicating
that C$_{6}$H$_{6}$ is more efficiently fragmented by soft X-ray
than UV radiation, where 50\% of the ionized benzene molecules
survive to UV dissociation while only about 4\% resist to X-rays.
Partial ion yields of H$^{+}$ and small hydrocarbons such as
C$_2$H$_2^{+}$, C$_3$H$_3^{+}$ C$_4$H$_2^{+}$ are determined as a
function of photon energy. Absolute photoionization and dissociative
photoionization cross sections have also been determined. From these
values, half-life of benzene molecule due to UV and X-ray photon
fluxes in CRL 618 were obtained.

\end{abstract}

\begin{keywords} 
astrochemistry -- methods: laboratory -- ISM: molecules -- X-rays:
stars -- molecular data --
\end{keywords}

\section{Introduction}
Benzene, C$_6$H$_6$, may be taken as the basic unit for the
polycyclic aromatic hydrocarbons (PAHs) which are believed to play
an important role in the chemistry of the interstellar medium (Woods
et al. 2003). It may also serve as a precursor molecule to more
complex organic compounds, such as amino acids like phenylalanine
and tyrosine.

It is known that polycyclic aromatic hydrocarbons (PAHs) are mainly
formed in the dust shells of late stages of asymptotic giant branch
type carbon rich stars (Cherchneff et al. 1992). Subsequently the
ejection into the interstellar medium of its C-rich envelope, these
stars become a proto-planetary nebulae (PPN) which evolve to
planetary nebulae (PN).

The PAHs are highly efficient absorbers of ultraviolet (UV)
radiation, which means they contribute considerably to interstellar
UV opacity (Draine 1978;  Bakes \& Tielens 1994). The UV radiation
absorbed by PAHs is converted from electronic to vibrational energy
resulting in infrared photon emission. The infrared spectra of
planetary nebulae show these strong emission features at 3.3, 6.2,
7.7 and 11.3 $\mu$m which have been attributed to aromatic
hydrocarbons and the features at 3.4 and 6.9 $\mu$m which might be
due to aliphatic hydrocarbons (Kwok et al. 2004). Shan et al. 1991
showed experimentally that the 3.4 - 3.6 $\mu$m band, observed in
some astronomical objects, may be assigned to stretching vibrational
excitation of CH$_3$ in methylated PAHs, or PAHs with a hydrogen
atom substituted by a methyl group. Vijh, Witt \& Gordon (2004)
evidenced the presence of the anthracene molecule (three benzenic
rings) through fluorescence spectra analysis in the PPN Red
Rectangle.

The detection of benzene, C$_{4}$H$_{2}$, C$_{6}$H$_{2}$, methyl
acetylene (CH$_{3}$C$_{2}$H) and methyl diacetylene
(CH$_{3}$C$_{4}$H) in the direction of the PPN CRL 618 was reported
by Cernicharo et al. (2001a; 2001b). All the infrared bands arose
from a region with kinetic temperatures between 200 and 250 K,
probably the photodissociation region (PDR) associated with the
material surrounding the central star. Small hydrocarbons have also
been detected in another similar object CRL 2688.

PAHs have also been observed within protostellar disks around both
low-mass T Tauri and intermediate-mass Herbig Ae/Be stars (Geers et
al. 2006, Acke \& van den Ancker 2004). T Tauri stars have strong
X-ray fields (Feigelson \& Montmerle 1999) that ionize the molecular
material at various distances around the protostellar producing a
complicated sequence of chemical reactions, mainly by ion-molecule
reactions.

The chemistry in circumstellar regions is strongly modified by the
UV photons emitted from the hot central star and by the X-rays
associated with its high-velocity winds. The soft X-rays are more
effective ionizing agents as compared to UV radiation because they
penetrate more deeply in the envelope and heat up the gas more
efficiently than UV (Deguchi et al. 1990). These authors have
suggested that the large observed abundance of the HCO$^{+}$ specie
can be explained by soft X-ray ionization rather than by UV
ionization. Using the Chandra X-ray observatory, Kastner et al.
(2003) have reported the discovery of X-ray emission from bipolar
planetary nebula.

Knowledge of photoabsorption, photoionization and
photo-fragmentation processes in the UV and X-rays regions for
knowing planetary nebulae is consequently extremely important. Le
Page et al. (2001) have pointed out on the important processes which may
affect the PAH distribution in the interstellar medium, such as
photo-fragmentation with carbon loss, double ionization, and
chemistry between PAH cations and minor species presented in
diffuse clouds.

The destruction of benzene isolated in two different matrices has
been measured for both UV photolysis and proton bombardment by
Ruiterkamp et al. (2005). They derived the destruction cross
sections for benzene and found that proton bombardment is more
efficient than the UV photolysis.

Using electron energy-loss spectroscopy, we have studied the
excitation and photoabsorption processes at the UV region (3-50 eV)
of benzene (Boechat-Roberty et al. 2004), naphthalene (de Souza et
al. 2002) and anthracene (Boechat-Roberty et al. 1997).

The present work is concerned with the experimental investigation of
the photoionization and dissociative photoionization of the benzene
molecule upon interaction with UV and soft X-ray (in the vicinity of
the carbon K shell), using synchrotron radiation and time of flight
mass spectrometry. In section 2, we present briefly the experimental
setup and techniques. The production of ions following the benzene
molecule dissociation by UV and soft X-ray photons, the
determination of the absolute cross sections and half-lives as a
function of photon fluxes are presented and discussed in section 3.
Finally, in section 4, final remarks and conclusions are given.

\section{Experimental}

The experiment was performed at the Brazilian Synchrotron Light
Laboratory (LNLS), S\~ao Paulo, Brazil. UV and soft X-rays photons
from a toroidal grating monochromator (TGM) beamline (12-310 eV),
perpendicularly intersect the gas sample inside a high vacuum
chamber. The gas needle was kept at ground potential. The emergent
photon beam flux ($\sim10^{12}$ photons s$^{-1}$) was recorded by a
light sensitive diode.
\begin{figure}
\resizebox{\hsize}{!}{\includegraphics{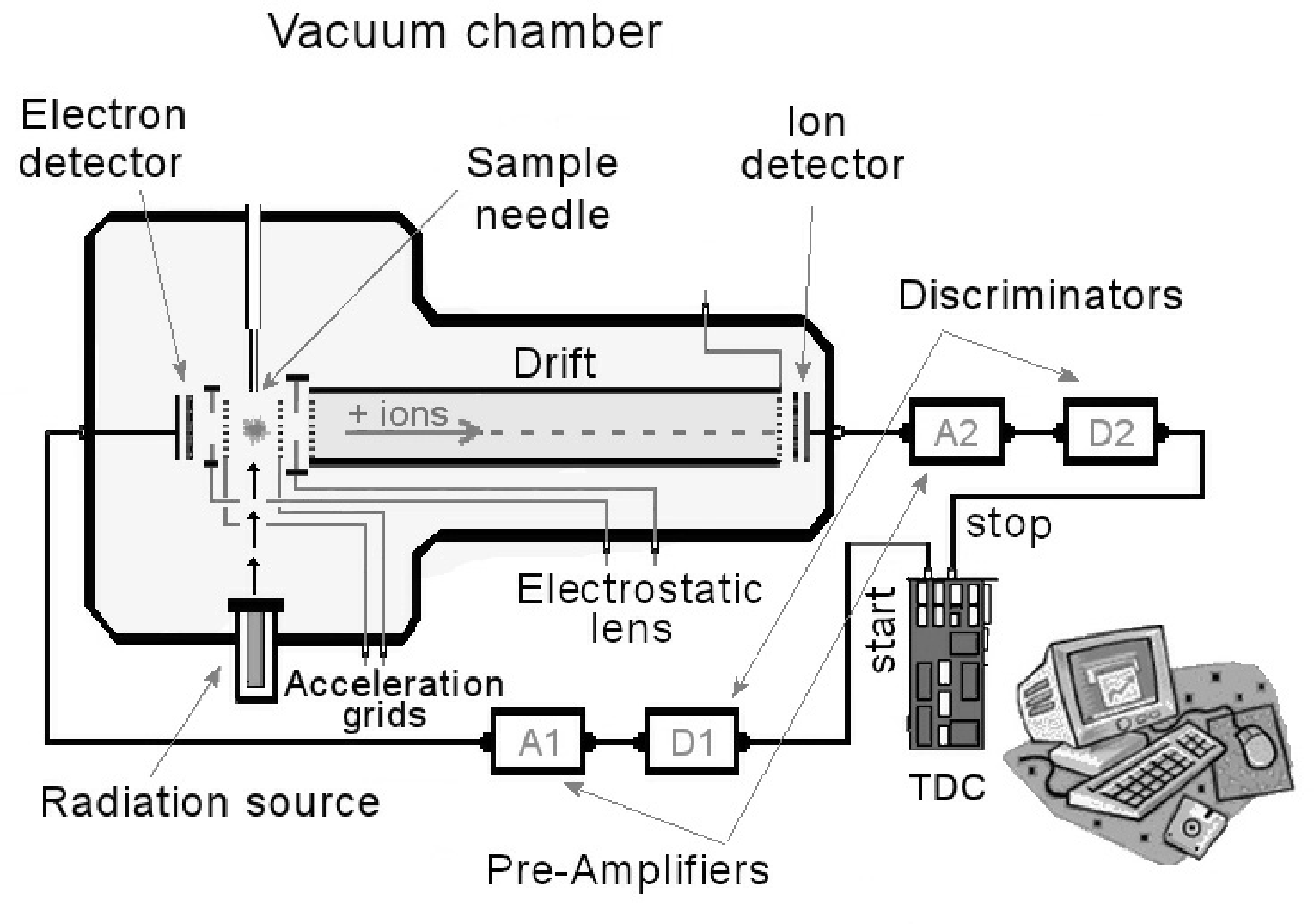}}
\caption{Schematic diagram of the time of flight mass spectrometer
(TOF-MS) inside the experimental vacuum chamber and the associated
electronics. See details in text.} \label{fig-diagram}
\end{figure}
Conventional time-of-flight mass spectra (TOF-MS) were obtained
using the correlation between one photoelectron and a photoion
(Photoelectron Photoion Coincidence PEPICO. The ionized recoil
fragments produced by the interaction with the photon beam were
accelerated by a two-stage electric field and detected by two
micro-channel plate detectors in a chevron configuration, after
mass-to-charge ($m/q$) analysis by a time-of-flight mass
spectrometer (297 mm long). They produced the stop signals to a
time-to-digital converter (TDC). Photoelectrons, accelerated in an
opposite direction with respect to the positive ions, are recorded
without energy analysis by two micro-channel plate detectors and
provide the start signal to the TDC. The first stage of the electric
field (708 V/cm) consists of a plate-grid system crossed at the
center by the photon beam. The TOF-MS was designed to have a
maximized efficiency for ions with kinetic energies up to 30 eV.
Negative ions may also be produced and detected, but the
corresponding cross-sections are negligible.

A schematic diagram of the time of flight spectrometer inside the
experimental vacuum chamber is shown in Fig.~\ref{fig-diagram},
where A1 and A2 are the pre-amplifiers and D1 and D2 are the
discriminators. The connection to the time-to-digital converter is
also shown. Besides PEPICO spectra, other two kinds of coincidence
mass spectra were obtained simultaneously, PE2PICO spectra
(PhotoElectron Photoion Photoion Coincidence) and PE3PICO spectra
(PhotoElectron Photoion Photoion Photoion Coincidence). These
spectra involve the detection of ions coming from double and triple
ionization processes, respectively, that arrive coincidentally with
photoelectrons. We have presented results on multi-ionization
followed dissociation of acetic acid and alcohols in Pilling et al.
(2007a; 2007b). Of all signals received by the detectors only about
10\% come from PE2PICO and 1\% from PE3PICO spectra, reflecting that
the major contribution is indeed due to single event coincidence.
Nonetheless, PEPICO, PE2PICO and PE3PICO signals were taken into
account for normalization purposes. Recoil ion and ejected electron
detection efficiencies of 0.23 and 0.04, respectively, were assumed.
In addition, we took 0.54 and 0.78 as the efficiencies to detect to
detect at least one of the photoelectrons from double ionization and
triple ionization events, respectively (Cardoso 2001). Benzene was
commercially obtained with high purity (99.7\%). No further
purification was performed other than degassing the liquid sample by
multiple freeze-pump-thaw cycles before admitting the vapor into the
chamber. Details on the time-of-flight spectrometer are available
elsewhere (Lago et al. 2004; Marinho et al. 2006).

The base pressure in the vacuum chamber was in the $10^{-8}$ Torr
range. During the experiment the chamber pressure was maintained
below $10^{-5}$ Torr. The pressure at the interaction region
(volume defined by the gas beam and the photon beam intersection)
was estimated to be $\sim$ 1 Torr (10$^{16}$ mols cm$^{-3}$). The
measurements were made at room temperature.

\section{Results and discussion}

\subsection{Production of ions}

\begin{figure}
\resizebox{\hsize}{!}{\includegraphics{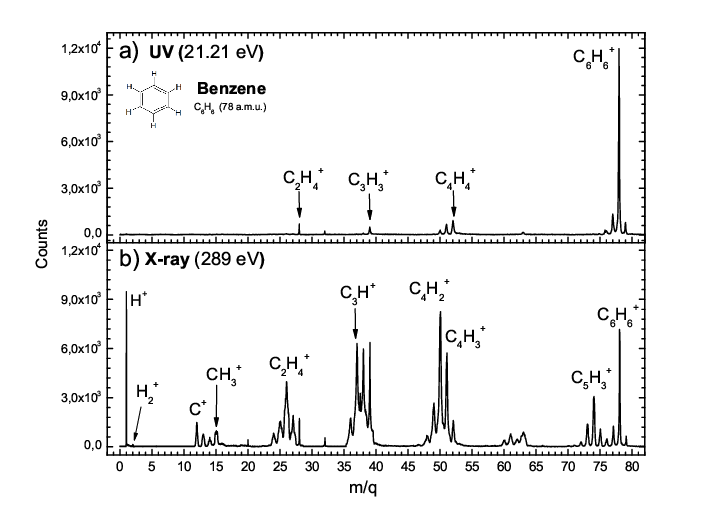}}
\caption{Time-of-flight mass spectra of benzene molecule recorded at
a) UV photons (21.21 eV) and b) soft X-ray photons (289 eV).}
\label{fig-count}
\end{figure}

Mass spectra of benzene were obtained at the UV energy of 21.21 eV,
that corresponds to the photon energy emitted by helium atom (HeI),
and at energies 282, 285, 289, and 301 eV, around the
C1s$\rightarrow\pi^{\ast}$ resonance at 285.2 eV (Hitchcock et al.
1987). The mass spectra obtained at 21.21 eV and 289 eV are shown in
Fig.~\ref{fig-count}. Clearly, X-rays produce much more types of
ions than UV photons. The most abundant ion at 21.21 eV corresponds
to the parent ion C$_{6}$H$_{6}^+$, confirming the relative high
stability of the benzene molecule at the UV energy range, whereas at
289 eV the molecule is highly destroyed giving rise to several
fragments like C$_4$H$_2^+$ (diacetylene) and to an enhancement of
the H$^+$ proton production.

The following ionic groups are due to carbon loss:  C group (from 12
to 15 amu), C$_2$ group (24 to 27 amu), C$_3$ group (36 to 39 amu),
C$_4$ group (48 to 52 amu), C$_5$ group (60 to 64 amu) and C$_6$
group (72 to 79 amu).

\begin{figure}
\resizebox{\hsize}{!}{\includegraphics{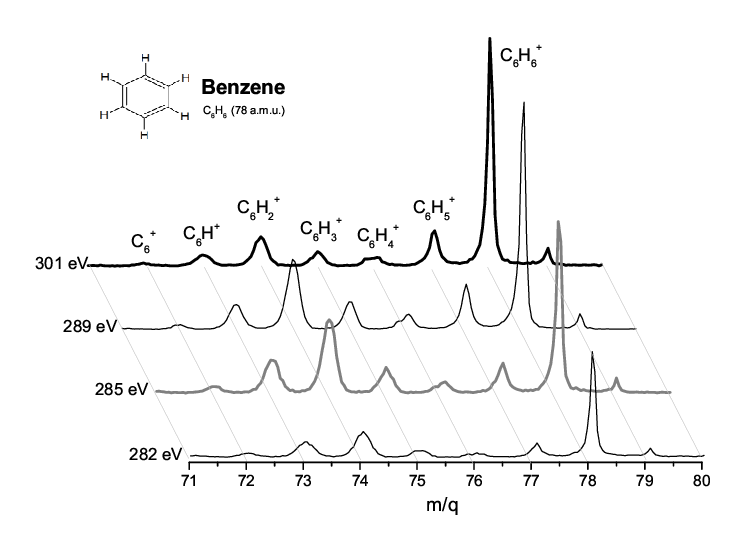}} \caption{Behavior
of C$_{6}$H$_{n}$ group in the fragmentation of benzene at energies
closer to the C1s resonance ($\sim$ 285 eV) as a function of the
energy.} \label{fig-benzene3d}
\end{figure}

In Fig.~\ref{fig-benzene3d} we present the fragmentation pattern due
to  multiple hydrogen loss from C$_{6}$H$_{6}$ as a function of the
photon energy. The phenyl radical, C$_{6}$H$_{5}$, derived from
benzene by removal of one hydrogen atom, is a possible progenitor of
other aromatic species. Kaiser et al. (2000) pointed out that
reactions of the phenyl radical with methylacetylene are relevant to
the chemistry of PAHs in extraterrestrial environments. They showed
that the reaction of C$_{6}$H$_{5}$ radicals with methylacetylene to
form phenylmethylacetylene is expected to play a role in PAH
synthesis only in high temperature interstellar environments, such
as circumstellar outflow of carbon star. The radical
C$_{6}$H$_{2}^{+}$ is the second most abundant ionic specie, in this
mass range, formed by hydrogen loss due to X-ray interaction and its
yield increases with the photon energy particularly near the C1s
resonance. Our data show that UV photons do not produce this radical
from benzene (or PAHs) dissociation and as the C$_{6}$H$_{2}$ was
observed in the CRL 618 (Cernicharo et al. 2001a), its abundance in
this environment might have a part due to the PAHs fragmentation by
X-rays.

\begin{figure}
\centering
\includegraphics[scale=0.65]{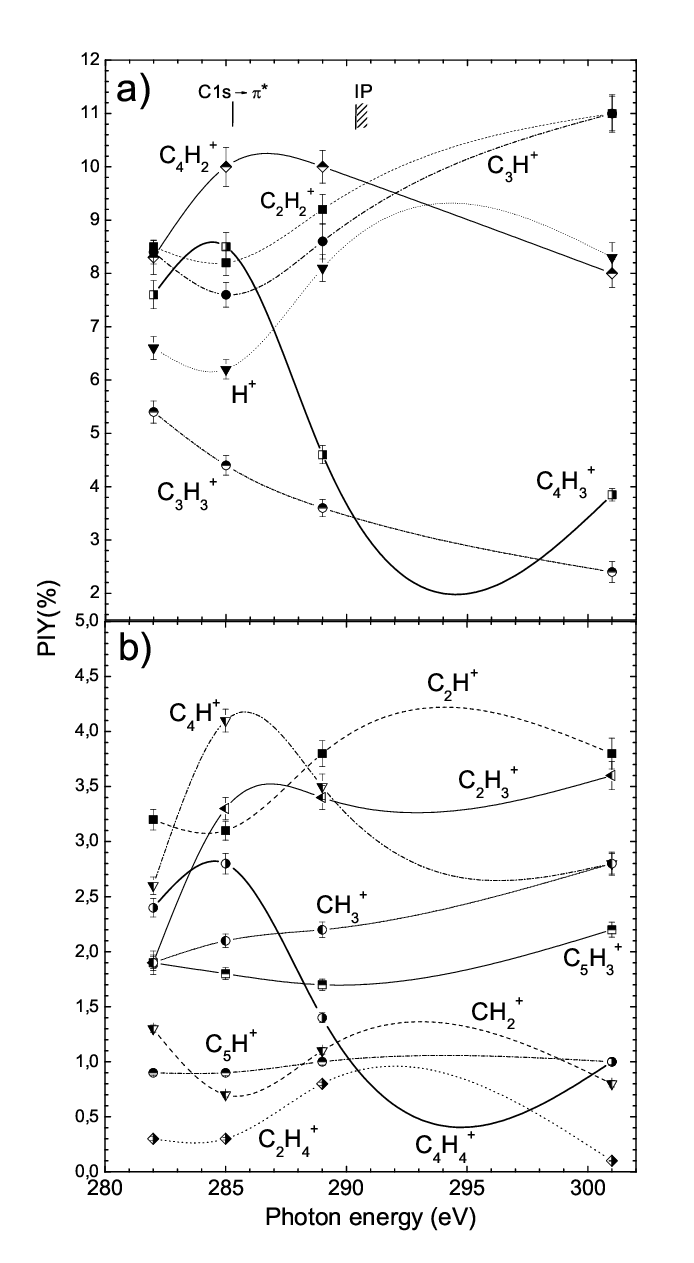} \caption{Partial ion yields (PIYs)
for some fragments of C$_{6}$H$_{6}$ molecule as a function of the
photon energy. a) PIY $>$ 5\% and b) PIY $<$ 5\%.} \label{fig-PIY}
\end{figure}
The partial ion yields (PIY) as a function of the photon energy are
shown in Fig.~\ref{fig-PIY} and Fig.~\ref{fig-PIY}b, for the most
significant outcomes (H$^+$, C$_{4}$H$_{2}^{+}$, C$_{2}$H$_{2}^{+}$,
C$_{3}$H$^{+}$, C$_{4}$H$_{3}^{+}$, C$_{3}$H$_{3}^{+}$) and for
other minor outcomes, respectively. The production of some ions like
C$_{4}$H$_{2}^+$ and C$_{4}$H$_{3}^+$ increases around the C1s
resonance (285.2 eV) and decreases above the ionization potential
(290.3 eV). However, the acetylene ion C$_{2}$H$_{2}^{+}$ shows an
inverse behavior. The statistical uncertainties are below 10\%.

The peak at $m/q$=39 amu may be assigned to C$_3$H$_{3}^{+}$ with
probably a contribution from C$_6$H$_{6}^{++}$. Braitbart et al.
1992 have obtained mass spectra for benzene in the 25 - 70 eV
region. They provided information on the relative yields of
C$_6$H$_{6}^{++}$/C$_3$H$_{3}^{+}$  and
C$_6$H$_{6}^{++}$/C$_6$H$_{6}^{+}$ at 35 and 70 eV. The
photoionization mass spectrum of benzene at 21.21 eV was also
discussed by Berkowitz (1979), where he presented relative intensity
values for ionic fragments taken from three different measurements.
As a example, the obtained average value of the ratio
C$_6$H$_{6}^{+}$/C$_3$H$_{3}^{+}$ is 9.13, in reasonable agreement
with our result, 10.0, taking into  account the present experimental
error which is about 10 \%. A comparison between our results with
Berkowitz (1979) and Braitbart et al. 1992 data is presented in Fig.
5a. The ratio  C$_6$H$_{6}^{+}$/C$_3$H$_{3}^{+}$ is seen to decrease
with increasing energy in the UV and soft X-Ray ranges. The
C$_6$H$_{6}^{++}$/C$_3$H$_{3}^{+}$ yield ratio is shown in the Fig.
5b. A value of about 0.17 was estimated at 21.21 eV by linear
fitting. It should however be taken into consideration that the peak
intensity at $m/q$=39 amu contains a 17\% contribution from benzene
dications. The formation process of benzene dications at energies
below the double  ionization potential (IP$^{++}$ = 26.1 eV)
proceeds by two stages (Leach 1995), such as C$_6$H$_{6}$ +
h$\nu_{1}{\rightarrow}$ C$_6$H$_{6}^{+}$ + e$^{-}$ and
C$_6$H$_{6}^{+}$ + h$\nu_{2}{\rightarrow}$ C$_6$H$_{6}^{++}$ +
e$^{-}$.

\begin{figure}
\centering
\includegraphics[scale=0.35]{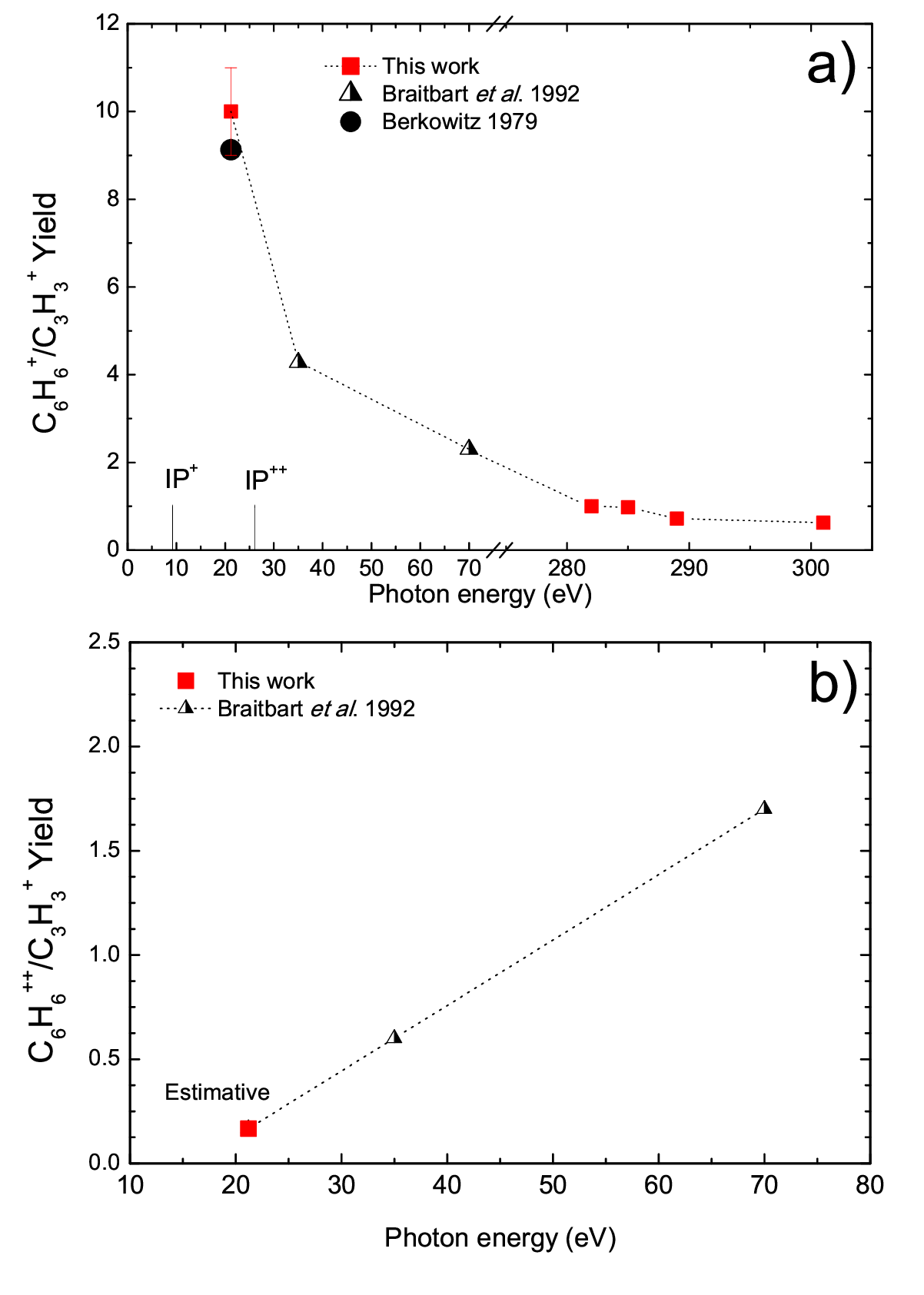}\caption{C$_{6}$H$_{6}^{+}$ and C$_3$H$_{3}^{+}$ yield
ratio as a function
of the photon energy. IP$^{+}$ and IP$^{++}$ are the single and
double ionization potentials, respectively.} \label{fig-ratio}
\end{figure}

The fragmentation pathways of C$_6$H$_6^{++}$ were calculated using
density functional theory by Rosi et al. (2004). They suggest that
Coulomb explosion is not a significant channel under most
astrophysical conditions and therefore, the dications are long-lived
and should be considered in the modeling of the interstellar medium.
The behaviour of PAH dications in HI and HII regions have been
discussed by Leach (1995) based on experimental results. It was
concluded that in HI regions, dication fragmentation will proceed
mainly by the covalent channel (A$^{++}{\longrightarrow}$  B$^{++}$
+ C) giving rise to small neutral products such as H, H$_2$,
C$_2$H$_2$, C$_2$H$_3$, C$_4$H$_2$. In HII regions, where more
energetic photons are available, the PAHs could be destroyed by
charge separation processes (A$^{++}{\longrightarrow}$  B$^{+}$ +
C$^{+}$) producing more ions like C$_3$H$_{3}^{+}$ and
C$_2$H$_{2}^{+}$. These ionic species could participate in the
formation of new molecules. Photodestruction of PAHs has been
proposed to explain the abundance enhancement of C$_3$H$_2$  or the
presence of small hydrocarbons in PDRs (Pety et al. 2005; Fuente et
al. 2003).

\begin{figure}
\resizebox{\hsize}{!}{\includegraphics{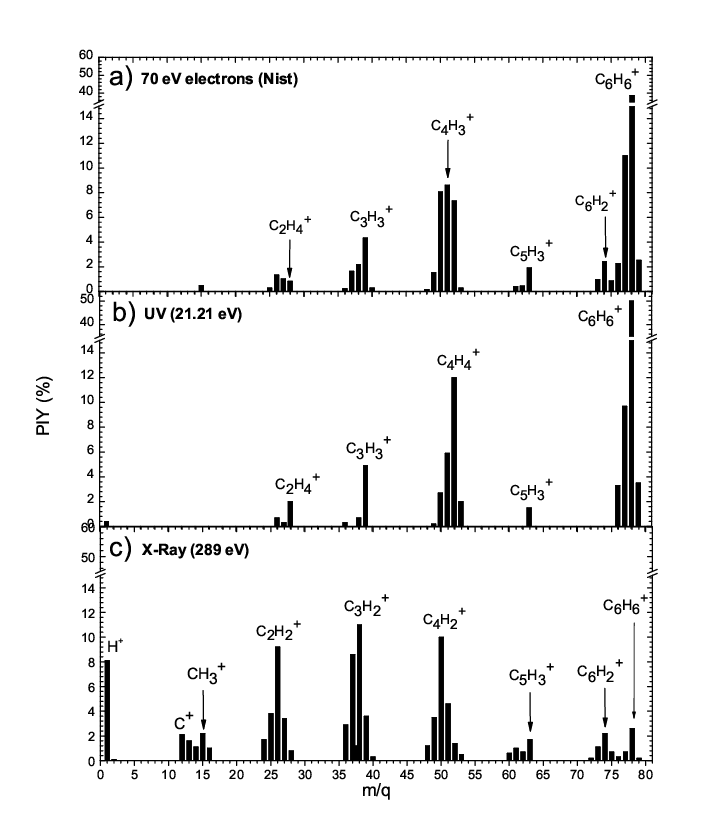}}\caption{Comparison
between partial ion yield (PIY) of C$_{6}$H$_{6}$ molecule obtained
with a) 70 eV electrons from NIST, b) 21.21 eV photons and c) 289 eV
photons.} \label{fig-nistlns}
\end{figure}

The partial ion yields (PIY) or relative intensities for the ionic
fragments observed in the C$_{6}$H$_{6}$ mass spectra recorded with
70 eV electrons (NIST data base), ultraviolet (21.21 eV) and soft
X-ray (289 eV) photons are compared in Figs.~\ref{fig-nistlns}a-c,
respectively. A comparison between the mass spectra recorded with 70
eV electrons and 21.21 eV photons reveals a similar fragmentation
pattern despite the different dissociation mechanisms. This
corroborate the idea that 70 eV electrons produce, as a first
approximation, the same ionic dissociation as created by ultraviolet
photons, as suggested before (Lago 2004; Boechat-Roberty 2005;
Pilling et al. 2006). The molecular ion C$_6$H$_6^+$ is clearly more
destroyed by soft X-rays than by UV photons, or only 3\% survive to
x-rays while 50\% resist to UV, as expected.

The major dissociative photoionization pathways (PIY $>$ 5\%) of
benzene due to ultraviolet (21.21 eV) and soft X-rays (282-301 eV)
photons are presented in tables~\ref{tab-pathUV} and
\ref{tab-pathXRAY}, respectively. We show that 12\% of C$_6$H$_6$
are UV dissociated via acetylene loss C$_2$H$_2^+$ while 9\% are
x-rays dissociated by releasing the ionized acetylene. The
dissociation channel with the liberation of the C$_4$H$_3^+$ +
neutrals represent about 5.9\% on both radiation regimens.

\begin{table*}

\centering \caption{Production of ions:
Partial Ion Yield - PIY (\%) as a function of photon energy. Only
fragments with intensity $>$ 0.1 \% were tabulated. The estimated
experimental error was below 10\%.} \label{tab-piy}
\begin{tabular}{ l l l r r r r r }
\hline \hline
\multicolumn{2}{c}{Fragments}    &  & \multicolumn{5}{c}{PIY (\%)}\\
\cline{1-2}  \cline{4-8}
   $m/q$        & Attribution    &  & 21.21 eV      & 282 eV    & 285 eV  & 289 eV     & 301 eV  \\
\hline
1       & H$^+$                  &  & 0.4           & 6.6       & 6.2      & 8.1        & 8.3\\
2       & H$_2^+$                &  & -             & 0.1      & 0.1    & 0.1        & 0.1\\
12      & C$^+$                  &  & -             & 1.8       & 1.5    & 2.1       & 2.1\\
13      & CH$^+$                 &  & -             & 1.5       & 1.2    & 1.6    & 1.7\\
14      & CH$_2^+$               &  & -             & 1.3       & 0.7    & 1.1    & 0.8\\
15      & CH$_3^+$               &  & -             & 1.9       & 2.1    & 2.2    & 2.8\\
16      & CH$_4^+$               &  & -             & 1.1       & 0.6    & 1.0    & 0.4\\
24      & C$_2^+$                &  & -             & 1.4      & 1.3     & 1.7    & 1.6\\
25      & C$_2$H$^{+}$           &  & -             & 3.2       & 3.1    & 3.8    & 3.8\\
26      & C$_2$H$_2^{+}$         &  & 0.7           & 8.5    & 8.2    & 9.2    & 11\\
27      & C$_2$H$_3^{+}$         &  & 0.3           & 1.9   & 3.3    & 3.4    & 3.6 \\
28      & C$_{2}$H$_{4}^+$       &  & 2.0           & 0.3    & 0.3   & 0.8   & 0.1 \\
36      & C$_3^+$                &  & -             & 2.9    & 2.1    & 2.9    & 2.9\\
37      & C$_3$H$^{+}$           &  & 0.3           & 8.4    & 7.6    & 8.6    & 11\\
38      & C$_3$H$_2^{+}$         &  & 0.7           & 11    & 6.3    & 11    & 15\\
39      & C$_{3}$H$_{3}^{+}$ (C$_{6}$H$_{6}^{++}$)  &  & 5.0  & 5.4    & 4.4   & 3.6   & 2.4\\
40      & C$_3$H$_4^+$           &  & -             & 0.8    & 0.3    & 0.3    & 0.4\\
48      & C$_{4}^+$              &  & -             & 0.9    & 0.9    & 1.2    & 1.2\\
49      & C$_4$H$^{+}$           &  & 0.2          & 2.6    & 4.1    & 3.5    & 2.8 \\
50      & C$_{4}$H$_{2}^+$       &  & 2.7          & 8.3    & 10     & 10    & 8.0 \\
51      & C$_{4}$H$_{3}^+$       &  & 5.9         & 7.6    & 8.5   & 4.6   & 3.2 \\
52      & C$_{4}$H$_{4}^+$       &  & 12           & 2.4   & 2.8   & 1.4    & 1.0 \\
53      & C$_4$H$_5^+$           &  & 2.0            & 0.8    & 0.7   & 0.5    & 0.4\\
60      & C$_{5}^+$               &  & -           & 0.5    & 0.5    & 0.6    & 0.5 \\
61      & C$_5$H$^+$            &  & -           & 0.9    & 0.9   & 1.0   & 1.0 \\
62      & C$_5$H$_2^+$            &  & -           & 0.8    & 0.8    & 0.7    & 0.5\\
63      & C$_5$H$_3^+$            &  & 1.5        & 1.9    & 1.8    & 1.7   & 2.2\\
72      & C$_6^+$                 &  & -           & 0.1   & 0.2   & 0.2   & 0.1 \\
73      & C$_6$H$^+$             &  & -           & 0.8   & 1.0  & 1.1    & 0.5\\
74      & C$_6$H$_2^+$             &  & -          & 1.6   & 2.5   & 2.2   & 0.7 \\
75      & C$_6$H$_3^+$             &  & -          & 0.7   & 0.9   & 0.7   & 0.2 \\
76      & C$_6$H$_4^+$           &  & 3.3      & 0.5  & 0.6   & 0.3   & 0.1 4\\
77      & C$_6$H$_5^+$           &  & 9.7     & 1.4   & 1.2   & 0.7 1  & 0.3 \\
78      & C$_6$H$_6^+$           &  & 50     & 5.4  & 4.3   & 2.6 & 1.5\\
79      & $^{13}$CC$_5$H$_6^+$   &  & 3.5    & 0.5  & 0.2   & 0.2  & 0.1 \\
\hline \hline
\end{tabular}
\end{table*}

The abundance of benzene depends on both the production and the
destruction rates. Therefore, knowledge of the photodissociation
processes and its ion yields plays an essential role in interstellar
chemistry. This works point out to the importance of the ionic
species to the increase of molecular complexity. In PDRs many ions
can be produced by the photodissociation of large molecules. We have
previously suggested that the production of H$_3^+$ via the
dissociative photoionization of methylated organic molecules
(Pilling et. al 2007c). In this work we are quantifying the
production of ions by benzene fragmentation.

The formation mechanism of complex molecules, such as benzene, is
considered to involve additions of C$_{2}$H$_{2}$ or C$_{2}$H and
subsequent closure to form a ring (Cherchneff et al. 1992). These
mechanisms use a neutral-neutral chemistry and need very high
densities and temperatures in the vicinity of 1000 K to be
efficient. Woods et al. (2002) have found that the enhanced ionizing
radiation results in a large abundance of a complex ring molecule.
The benzene molecule can be formed efficiently near regions with
intense ionizing sources such as a high X-ray photons. The authors
have also shown that the physical conditions in CRL 618 may
efficiently contribute to the formation of C$_6$H$_6$. In addition,
the formation mechanism of benzene depends on the environment with
different pathways in planetary nebulae and in the interstellar
medium (Woods \& Willacy 2007).

\begin{table}
\caption{Main dissociation pathways from single ionization due to
ultraviolet photons (21.21 eV)} \label{tab-pathUV}
\setlength{\tabcolsep}{4pt}
\begin{center}
\begin{tabular}{l c l}
\hline \hline
C$_6$H$_6$ + $h\nu$ & $\stackrel{}{\longrightarrow}$ & C$_6$H$_6^+  + e^-$  \\
\hline
C$_6$H$_6^+$  & $\stackrel{12\%}{\longrightarrow}$  & C$_4$H$_4^+$ + C$_{2}$H$_{2}$ \\
              & $\stackrel{9.7\%}{\longrightarrow}$ & C$_6$H$_5^+$ + H  \\
              & $\stackrel{5.9\%}{\longrightarrow}$  & C$_4$H$_3^+$  + C$_{2}$H$_{2}$ + H \\
              & $\stackrel{5.0\%}{\longrightarrow}$  & C$_3$H$_3^+$  + neutrals  \\
\hline
\end{tabular}
\end{center}
\end{table}

\begin{table}
\caption{Main dissociation pathways from single ionization due to
soft X-rays (282-301 eV)} \label{tab-pathXRAY}
\setlength{\tabcolsep}{4pt}
\begin{center}
\begin{tabular}{l c l}
\hline \hline
C$_6$H$_6$ + $h\nu$ & $\longrightarrow$ & C$_6$H$_6^+  + e^-$    \\
\hline
C$_6$H$_6^+$  & $\stackrel{11\%}{\longrightarrow}$  & C$_3$H$_2^+$ + neutrals   \\
              & $\stackrel{9.2\%}{\longrightarrow}$  & C$_2$H$_2^+$ + neutrals    \\
              & $\stackrel{9.1\%}{\longrightarrow}$ & C$_4$H$_2^+$ + neutrals   \\
              & $\stackrel{8.9\%}{\longrightarrow}$  & C$_3$H$^+$  + neutrals   \\
              & $\stackrel{7.3\%}{\longrightarrow}$  & H$^+$  + neutrals        \\
              & $\stackrel{5.9\%}{\longrightarrow}$ & C$_4$H$_3^+$ + neutrals \\
\hline
\end{tabular}
\end{center}
\end{table}

\subsection{Half-lives of benzene and UV and X-ray photon fluxes}

Absolute cross section values for both photoionization and
photodissociation processes of organic molecules are extremely
important as inputs for molecular abundances models (Sorrell 2001).
In these theoretical models, molecules are formed inside the bulk of
icy grain mantles photoprocessed by starlight (ultraviolet and soft
X-ray photons). The main chemistry route is based on radical-radical
reactions followed by chemical explosion of the processed mantle
that ejects organic dust into the ambient gaseous medium. The
destruction of a given molecule subjected to a radiation field in
the energy range $E_{2} - E_{1}$ inside a gaseous-dusty cloud can be
written by (Cottin et al. 2003)
\begin{equation}
-\frac{dN}{dt}= N k_{ph}
\end{equation}
where ${N}$ is the column density ( molecules cm$^{-2}$) and
$k_{ph}$ is the photodissociation rate (s$^{-1}$) given by
\begin{equation} \label{eq-R}
k_{ph} = \int_{E_1 }^{E_2} \sigma_{ph-d}(E) f(E) dE
\end{equation}
where $\sigma_{ph-d}(E)$ (cm$^{2}$) is the photodissociation cross
section and $f(E)$ is the photon flux both as a function of energy
(photons cm$^{-2} eV^{-1} s^{-1}$). The photoionization rate is also
given by
\begin{equation} \label{eq-i}
\zeta_{i} = \int_{E_1 }^{E_2} \sigma_{ph-i}(E) f(E) dE
\end{equation}
where $\sigma_{ph-i}(E)$ (cm$^{2}$) is the photoionization cross
section.

The determination of $\sigma_{ph-d}$ of molecules is very important
to estimate the molecular abundance in the interstellar
environments. Moreover, knowing the photon dose and $\sigma_{ph-d}$
values its is possible to determine the half-life of a given
molecule.

The half-life may be obtained from the Eq.~\ref{eq-R} by writing
\begin{equation}
t_{1/2} = \frac{\ln2}{k_{ph}}
\end{equation}
which does not depend on the molecular number density.

\begin{figure}
\resizebox{\hsize}{!}{\includegraphics{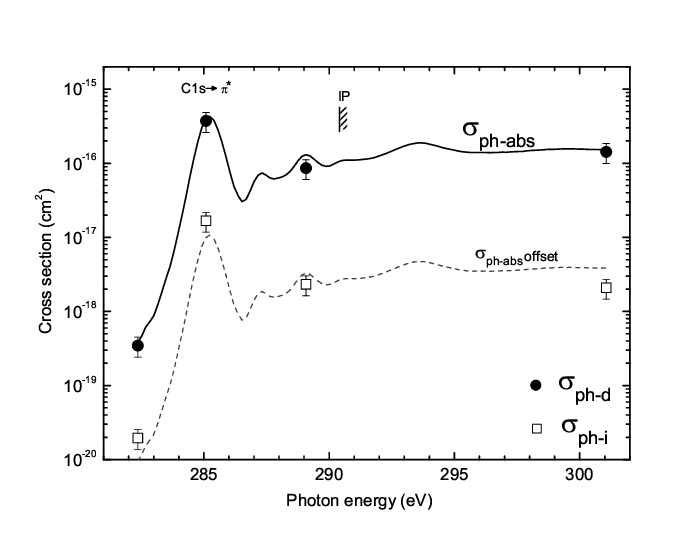}}
\caption{Non-dissociative single ionization (photoionization) cross
section, $\sigma_{ph-i}$ ({\tiny $\square$ }) and dissociative
ionization (photodissociation) cross section, $\sigma_{ph-d}$
($\bullet$), of benzene as a function of photon energy. The
photoabsorption cross-section, $\sigma_{ph-abs}$ (solid line), was
taken from Hitchcock et al. (1987) is also shown. The dashed line is
an off-set of photoabsorption cross-section, only to guide the
eyes.} \label{fig-secaochoque}
\end{figure}

Assuming a negligible fluorescence yield due to the low carbon
atomic number (Chen et al. 1981) and discarding anionic fragments in
the present X-ray photon energy range, we assumed that all absorbed
X-ray photons lead to cationic ionizing process. Therefore, the
precise determination of non-dissociative single ionization cross
section ($\sigma_{ph-i}$) and the dissociative single ionization
(photodissociation) cross section ($\sigma_{ph-d}$) of benzene can
be determined by
\begin{equation}
\sigma_{ph-i} = \sigma^{+} \frac{PIY_{C_6H_6^+}}{100}
\end{equation}
and
\begin{equation}
\sigma_{ph-d} = \sigma^{+} \Big( 1 - \frac{PIY_{C_6H_6^+}}{100}
\Big)
\end{equation}
where $\sigma^{+}$ is the cross section for single ionized fragments
(see description in Boechat-Roberty et al. 2005 and Pilling et al.
2006).

In order to put our data on an absolute scale, after subtraction of
the linear background and false coincidences coming from aborted
double and triple ionization (see Simon et al. 1991), we have summed
the contributions of all cationic fragments and the sum has been
normalized to the photoabsorption cross sections measured by
Hitchcock et al. (1987).

Absolute dissociative photoionization and photoionization cross
sections as a function of the X- ray photon energy can be seen in
Fig.~\ref{fig-secaochoque}. The absolute absorption cross section of
benzene (Hitchcock et al. 1987) is also shown for comparison (solid
line). These values are also shown in Table~\ref{tab-sigma}. The
$\sigma_{ph-d}$ is about 0.93 of $\sigma_{ph-abs}$ and
$\sigma_{ph-i}$ is about 0.033 of $\sigma_{ph-abs}$.

In the UV interstellar radiation field (ISRF) a benzene molecule is
destroyed by acetylene loss at a rate of 1.5 x 10$^{-10}$s$^{-1}$
(Woods \& Willacy 2007) which correspond to a survival time of
around 200 years. Taken into account that the diffuse interstellar
UV flux, integrated from 6 to 13.6 eV, is 10$^{8}$ photons cm$^{-2}$
s$^{-1}$, the integrated cross sections is $\sigma_{UV}=$ 1.5 x
10$^{-18}$ cm$^{2}$. However, Ruiterkamp et al. (2005) reported that
the half-life of gas phase benzene due to photolysis in diffuse
clouds is 27 years, that corresponding to a $\sigma_{UV}=$ 8.1 x
10$^{-18}$ cm$^{2}$. Therefore, this latter cross section is 5.4
times higher than the dissociation by acetylene loss only.

As mentioned before, employing electron energy-loss spectroscopy, we
have determined photoabsorption cross sections ($\sigma_{ph-abs}$)
at the UV region (3-45 eV) for the benzene molecule (Boechat-Roberty
et al. 2004). Integrating $\sigma_{abs}$(E) from 6 to 13.6 eV we
obtained 4.93 x 10$^{-16}$cm$^{-2}$, which corresponds to 61.1 times
$\sigma_{UV}$, or

$\int_{6eV}^{13.6eV}\sigma_{ph-abs}dE$ = 61.1
$\int_{6eV}^{13.6eV}\sigma_{ph-d}dE$ = 61.1 $\sigma_{UV}$.

For the X-ray range, integrating the our dissociative
photoionization cross section values from 280 to 310 eV
\begin{equation} \label{eq-cs}
\sigma_{X} = \int_{280 eV }^{310 eV} \sigma_{ph-d}(E)dE
\end{equation}
we obtained the value of 3.8 x $10^{-17}$ cm$^{2}$. Integrating the
photoionization cross sections values at the same range we found 1.2
x $10^{-18}$ cm$^{2}$.

\begin{figure}
\resizebox{\hsize}{!}{\includegraphics{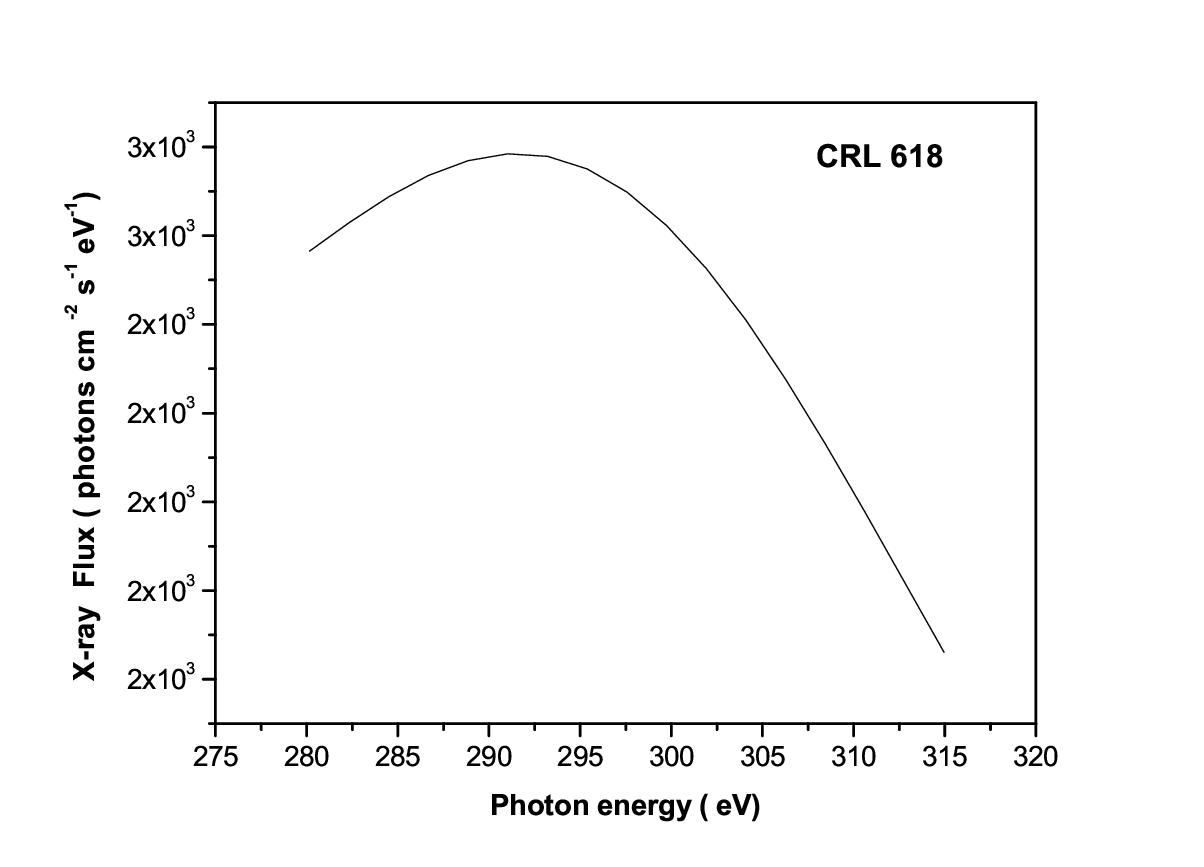}} \caption{CRL 618
X-ray spectrum at 8.9 x 10$^{15}$ cm (595 AU) from the central star
and assuming that it is 1500 pc away. Adapted from Lee \& Sahai
(2003) model flux.} \label{fig-CRL618flux}
\end{figure}

In spite of the fact that X-rays have not been detected in CRL 618
yet (Guerrero, Chu, \& Gruendl 2000, Guerrero et al. 2006), for
future observations, Lee \& Sahai (2003) proposed a X-ray emission
spectrum for CRL 618 in the 0.2 - 1.5 keV energy range derived from
a model flux. This spectrum was used to estimate the soft X-ray
photon flux at 280-310 eV range  impinging on benzene molecules
located at about 8.9 x 10$^{15}$cm (595 AU) from the central star,
where the majority of species are more abundant(Woods et al. 2003).
It was also considered that CRL 618 is distant 1500 pc from the
Earth (Fig.~\ref{fig-CRL618flux}). The integrated photon flux from
280 to 310 eV

\begin{equation} \label{eq-flux}
F_{X} = \int_{280 eV }^{310 eV} f(E)dE
\end{equation}
is equal to 7.6 x 10$^{4}$ (photons cm$^{-2}$ s$^{-1}$) which
corresponds to a half-life of the benzene of 7.6 x 10$^{3}$ years.
As the X-rays extinction was not taken into account in this
calculation, the actual photon flux will be consequently smaller
than the presented value. For comparison, approximately the same
flux is present at the dense molecular cloud AFGL 2591   (Stauber et
al. 2005). The photon flux (280-310 eV) from the Sun at 1 UA is
about 8 x 10$^{7}$ photons cm$^{-2}$ s$^{-1}$ (Gueymard 2004) which
the half-life is only 7.2 years.

\begin{figure}
\resizebox{\hsize}{!}{\includegraphics{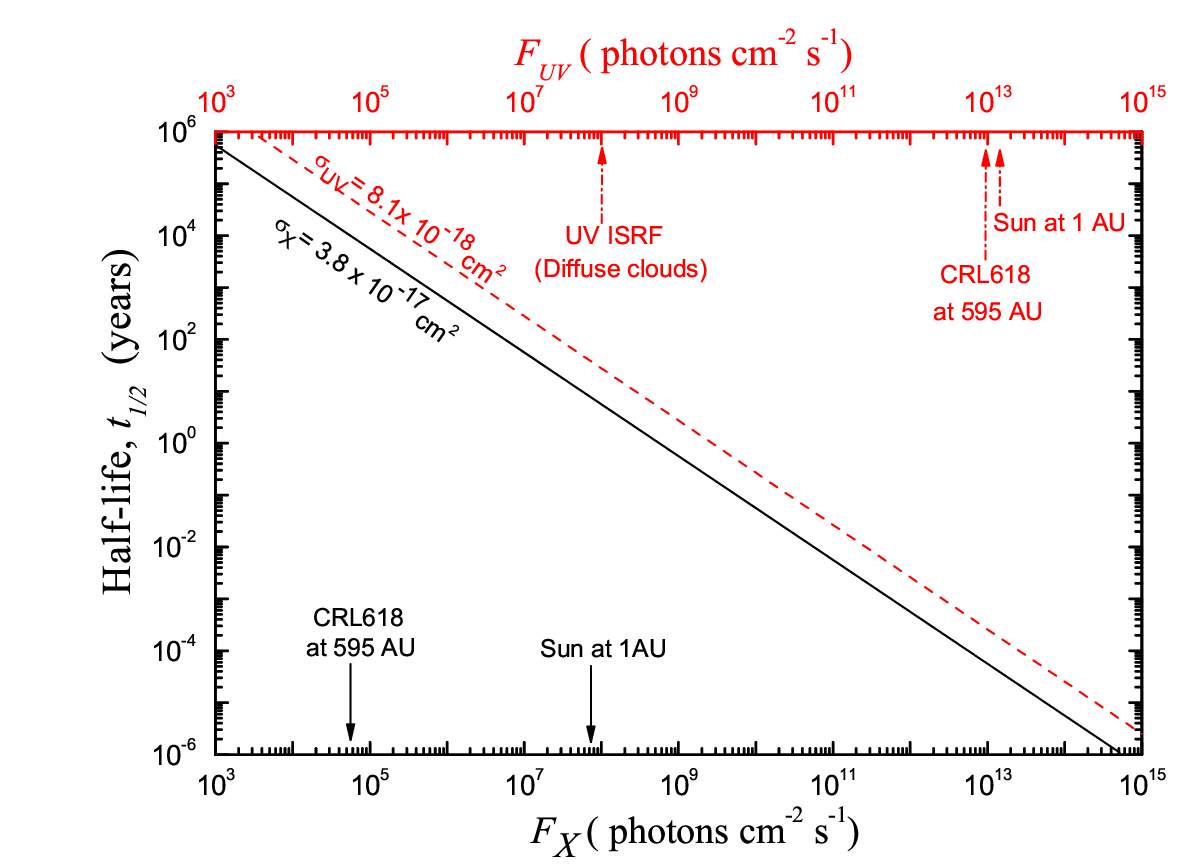}}
\caption{Half-life of the benzene molecule as a function of UV and
X-ray integrated photon fluxes, F$_{UV}$ (from 6 to 13.6 eV) and
F$_{X}$ (from 280 to 310 eV), respectively.} \label{fig-photonflux}
\end{figure}

The half-life of the benzene molecule as a function of UV and X-ray
photon fluxes is shown in the Fig.~\ref{fig-photonflux}. The cross
section $\sigma_{X}$ is 4.7 times higher than the $\sigma_{UV}$.
However, in general, the stellar UV photon flux is higher than X-ray
one. For the  Sun, F$_{UV}$ is about 10$^{4}$ F$_{X}$. In the CRL
618, the UV radiation field emitted from the central star B0-O9.5 is
about 2 x 10$^{5}$ times the ISRF, (2 x 10$^{13}$ photons cm$^{-2}$
s$^{-1}$), at a distance of 1 x 10$^{16}$ cm, or 668.5 AU (Herpin \&
Cernicharo 2000), leading to a half-life of about 2 hours. The
molecular dissociation due to X-rays becomes more important than by
UV mainly in X-ray dominate regions, XDR, where UV radiation are
almost totally attenuated.

\begin{table}
\centering \caption{Values of non-dissociative single ionization
(photoionization) cross section ($\sigma_{ph-i}$) and dissociative
ionization (photodissociation) cross section ($\sigma_{ph-d}$) of
benzene as a function of photon energy. The estimated total error is
30\%. The photoabsorption cross section ($\sigma_{ph-abs}$) from
Hitchcock et al. (1987) is also shown.} \label{tab-sigma}
\begin{tabular}{ l l c c c }
\hline \hline
Photon       &  & \multicolumn{3}{c}{Cross Sections (cm$^{2}$)}\\
\cline{3-5} energy (eV)  &  & $\sigma_{ph-d}$ & $\sigma_{ph-i}$     & $\sigma_{ph-abs}$ \\
\hline
282          &  & 3.5 $\times 10^{-21}$  & 1.9 $\times 10^{-22}$  & 3.8 $\times 10^{-21}$ \\
285          &  & 3.7 $\times 10^{-18}$  & 1.7 $\times 10^{-19}$  & 4.2 $\times 10^{-18}$ \\
289          &  & 9.7 $\times 10^{-18}$  & 2.6 $\times 10^{-20}$  & 1.3 $\times 10^{-18}$ \\
301          &  & 1.3 $\times 10^{-18}$  & 2.1 $\times 10^{-20}$  & 1.5 $\times 10^{-18}$ \\
\hline \hline
\end{tabular}
\end{table}


\section{Conclusions}

We have performed an experimental study of the photoionization and
dissociative photoionization processes of the benzene molecule using
synchrotron radiation and time-of-flight mass spectrometry in the
electron-ion coincidence mode.

It was observed that the benzene molecules are more efficiently
fragmented by soft X-rays, producing many more ions, than UV
photons.  50\% of the C$_6$H$_6^{+}$ molecules survive to UV
dissociation while only about 4\% resist to X-ray destruction.
Partial ion yields of H$^{+}$ and small hydrocarbons such as
C$_4$H$_2^{+}$, C$_2$H$_2^{+}$, C$_3$H$_3^{+}$  are determined as a
function of photon energy.

The abundance of the hydrocarbons C$_4$H$_2$, C$_6$H$_2$,
C$_2$H$_4$, CH$_3$C$_2$H, CH$_3$C$_4$H detected in the CRL 618,
could also be associated with the photodissociation of benzene, PAHs
and their methyl derivatives that after might undergo the radiative
recombination.

The X-ray dissociative photoionization cross section $\sigma_{X}$
(integrated in the 280 - 310 eV range) is 4.7 times higher than the
UV photodissociation cross section $\sigma_{UV}$ (integrated from 6
to 13.6 eV). As the stellar UV photon flux is more intense than
X-ray one, UV destruction rate is larger than X-ray destruction
rate, despite X-rays produce many more types of ionic fragments.

From the absolute UV and X-ray photodissociation cross sections,
$\sigma_{UV}$ and $\sigma_{X}$ for the benzene molecule it was
possible to obtained the half-life as a function of photon fluxes,
that can be used to know about the survival of this molecule in
circumstellar and interstellar environments where the PAHs are
present.

%
\section*{Acknowledgments}
The authors would like to thank the staff of the Brazilian
Synchrotron Light Laboratory (LNLS) for their valuable help during
the experiments. We are particularly grateful to Professor A.
Naves de Brito for the use of the Time-of-Flight Mass
Spectrometer. This work was supported by LNLS, CNPq, CAPES, FAPESP
and FAPERJ.

%
%

\bsp

\label{lastpage}

\end{document}